\documentclass[aps]{revtex4}
\usepackage{graphicx,color,psfrag}
\def\bc{\begin{center}}
\def\ec{\end{center}}

\def\beq{\begin{equation}}
\def\eeq{\end{equation}}

\def\d{\downarrow}
\def\u{\uparrow}

\begin{document}

%e-mail: ziegler@physik.uni-augsburg.de\\
%telephone: (49) 821 598 3244, FAX: (49) 821 598 3262

%\maketitle
%Title of paper
\title{Robust optical conductivity in gapped graphene 
\\
%{\small (ocond.tex)}
}

\author{K. Ziegler and A. Sinner}
\affiliation{
Institut f\"ur Physik, Universit\"at Augsburg\\
D-86135 Augsburg, Germany
}
\date{\today}

\begin{abstract}
We study the optical conductivity in the low-energy regime of gapped
mono- and bilayer graphene. A scaling relation is found, in which the
four parameters frequency, gap, Fermi energy and temperature appear only
as combination of three independent parameters. The ratio of the optical
conductivity of bilayer and mononlayer graphene is exactly 2.
\end{abstract}


\maketitle

\section{Introduction}

Monolayer and bilayer graphene are semimetals with good conducting
properties \cite{novoselov05,zhang05,geim07}. 
Moreover, in the presence of a microwave field the related optical
conductivity is constant over a large regime of frequencies. This has
been found in theoretical calculations 
\cite{gusynin07,ziegler07,mikhailov07,stauber08,gusynin09}
and was also observed experimentally \cite{nair08,basov08}. The reason for this behavior
is the existence of at least two bands in both materials, where at Fermi
energy $E_F=0$ (i.e. graphene without a gate potential) the lower band
is occupied and the upper band is unoccupied. As consequence, the absorption
of photons of energy $\hbar\omega$ from the microwave field creates 
electron-holes pairs due to the excitation of electrons from the 
lower band at energy $-\hbar\omega/2$ to the unoccupied upper band 
at energy $\hbar\omega/2$. This mechanism applies also to gated graphene
which has a shifted Fermi energy $E_F\ne0$. However, in this case photons can
only be absorbed if $\hbar\omega/2> E_F$ (for $E_F>0$), since all the states
in the upper band are occupied up to the energy $E_F$. Correspondingly, a photon
can only be absorbed for $-\hbar\omega/2< E_F$ if $E_F<0$. This means that
electron-hole creation by a microwave field is only possible if $\hbar\omega>|E_F|$

\begin{figure}[t]
\psfrag{E}{$E$}
\psfrag{k1}{$k_1$}
\psfrag{om}{$\hbar\omega$}
\begin{center}
%\psfrag{H2}{${\cal H}_2$}
%\psfrag{H4}{${\cal H}_4$}
\includegraphics[width=5cm,height=5cm]{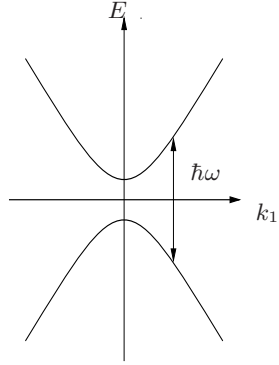}
\caption{Schematic picture of the creation of an electron-hole pair 
in gapped mono- or bilayer
graphene by the absorption of a photon with energy $\hbar\omega$.
For this process the photon energy must be larger than the band gap $\Delta=2m$. 
}
\label{paircreation}
\end{center}
\end{figure}

It has been found in a number of recent experiments that the creation of a gap 
in the semimetallic band structure of monolayer graphene (MLG) is possible by
absorption of hydrogen \cite{elias09} or in bilayer graphene (BLG) by applying a double
gate \cite{ohta06}. In both cases an electron-holes pair can also be created but this
requires a photon energy larger than the band gap $\Delta$ (cf. Fig. \ref{paircreation}).

Once electron-holes pairs have been created they will contribute to a current in
the material, where the latter is related to the strength of the external microwave
field by the optical conductivity $\sigma_{\mu\mu}(\omega)$. This quantity can
be measured experimentally and characterizes the electronic properties of the material.
In particular, it can be used to determine the band gap $\Delta$, since it vanishes for 
$\hbar\omega\le \Delta$. BLG, in contrast to MLG, has two low- and two high-energy bands.
As a result, there are several gaps that lead to electron-hole pair creations on different
energy scales with a more complex behavior of the optical conductivity 
\cite{abergel07,nilsson08,nicol08}.

In the following the optical conductivity shall be evaluated via the Kubo formalism
for the low-energy bands in MLG and BLG at nonzero temperature $T$. This avoids electron-hole
pair creation from higher energy bands and van Hove singularities. An important
question in this context is the role of the low-energy quasiparticle spectrum on
the optical conductivity. In order to focus on simple spectral properties, we consider only 
non-interacting electrons in a periodic tight-binding model.
Thus disorder, electron-electron interaction and electron-phonon interaction
are not taken into account.

\section{Model calculation}

The low-energy quasiparticle states in MLG with a gap $\Delta=2m$
are described by the massive two-dimensional Dirac equation
\beq
{\bf H}_M\pmatrix{
\Psi_\u \cr
\Psi_\d \cr
}=\pmatrix{ 
-E+m& i\partial_x+\partial_y \cr
i\partial_x-\partial_y & -E -m\cr
}\pmatrix{
\Psi_\u \cr
\Psi_\d \cr
}=0
\ .
\label{diracequ00}
\eeq
For simplicity, we have set the Fermi velocity $v_F=1$ because this
parameter will not appear in the final results for the conductivity. 
A similar equation exists for the low-energy quasiparticle states of BLG 
\cite{katsnelson07,castro08}:
\beq
{\bf H}_B\pmatrix{
\Psi_\u \cr
\Psi_\d \cr
}=\pmatrix{ 
-E+m& (i\partial_x+\partial_y)^2 \cr
(i\partial_x-\partial_y)^2 & -E -m\cr
}\pmatrix{
\Psi_\u \cr
\Psi_\d \cr
}=0
\ .
\label{diracequ01}
\eeq
With the plane-wave ansatz $\Psi_\d(x,y)=\Psi_\d e^{ik_xx+ik_yy}$ we obtain for MLG
the following relations
\beq
\Psi_\d(x,y)=\frac{k_x+ik_y}{E+m}\Psi_\u(x,y) ,\ \ E^2=m^2+k^2
\label{eigen1}
\eeq
and for BLG
\beq
\Psi_\d(x,y)=\frac{(k_x+ik_y)^2}{m+E}\Psi_\u(x,y) ,\ \ 
E^2=m^2+k^4
\ .
\label{eigen2}
\eeq
These solutions will be used as a basis for evaluating current matrix elements
and the optical conductivity.

%\section{Kubo formula}

{\it Kubo formula:}
The optical conductivity can be calculated from the Kubo formula. 
This requires the evaluation of the current operator $j_\mu=ie[{\bf H},r_\mu]$,
where $r_\mu$ ($\mu=1,2$) is a component of the position of the quasiparticle.
The nonzero matrix elements of the current operator with respect to the energy eigenstates of 
Eqs. (\ref{eigen1}), (\ref{eigen2}) are either diagonal elements 
$ie\langle E_k|[H,r_\mu]|E_k\rangle$ or the off-diagonal elements
$ie\langle E_k|[H,r_\mu]|-E_k\rangle$. It turns out that the diagonal elements
do not appear in the real part of the optical conductivity \cite{ziegler06}.
Only the off-diagonal terms contribute because the optical conductivity requires a 
scattering process between two states whose energy difference is just
the photon energy $\omega$. A convenient representation of the Kubo formula then is \cite{ziegler06}
\[
\sigma_{\mu\mu}'\equiv Re(\sigma_{\mu\mu})=\frac{\pi e^2}{4\hbar}\int
|\langle E_k|[H,r_\mu]|-E_k\rangle|^2\delta(E_k+\omega/2)
\frac{f_\beta(E_F-E_k)-f_\beta(E_F+E_k)}{E_k}\frac{d^2k}{(2\pi)^2}
\]
with the Dirac-Fermi distribution $f_\beta(x)=1/(1+e^{\beta x})$
at inverse temperature $\beta=1/k_B T$. (Here and in the rest of this paper
the spin and valley degeneracy, providing an extra factor 4, has not been written explicitly.) 
Integration over $E_k$ gives
\beq
\sigma_{\mu\mu}'=-\frac{e^2}{4h}\int_0^{2\pi}
|\langle \omega/2|[H,r_\mu]|-\omega/2\rangle|^2d\varphi 
J
%k\left(\frac{dE_k}{dk}\right)^{-1}
\frac{f_\beta(E_F+\omega/2)-f_\beta(E_F-\omega/2)}{\omega}
\label{kubo0}
\eeq
with the Jacobian $J$ from the radial $k$ integration:
\[
%k\left(\frac{dE_k}{dk}\right)^{-1}
J=\Theta(\omega^2-m^2/4)\cases{
\omega/2& MLG \cr
\omega/2\sqrt{\omega^2-4m^2} & BLG \cr
}
\ .
\]

%\section{Current matrix elements}

\begin{figure}[t]
\psfrag{x}{$\varphi$}
\psfrag{y}{current}
\begin{center}
%\psfrag{H2}{${\cal H}_2$}
%\psfrag{H4}{${\cal H}_4$}
\includegraphics[width=7cm,height=7cm]{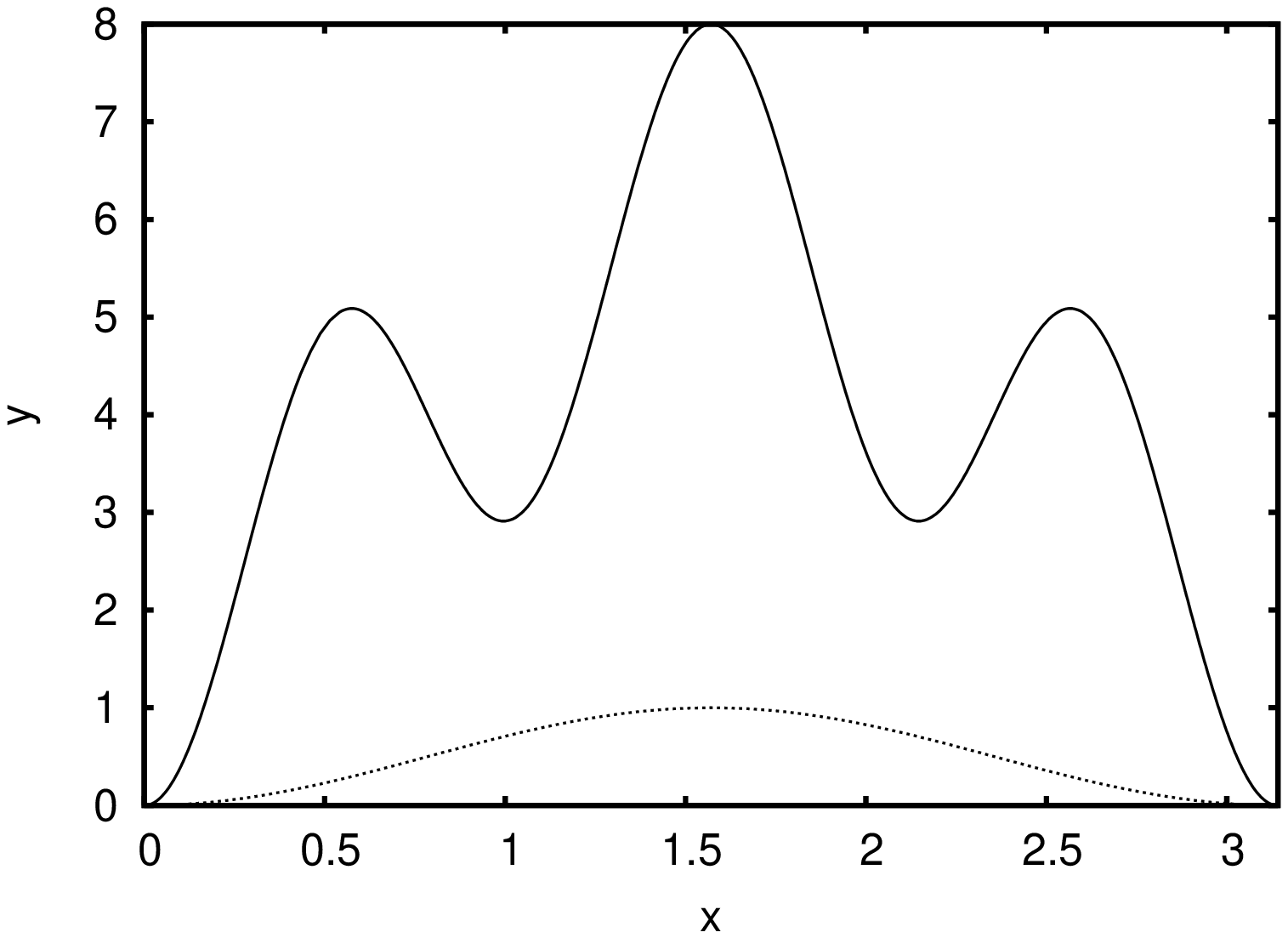}
\includegraphics[width=7cm,height=7cm]{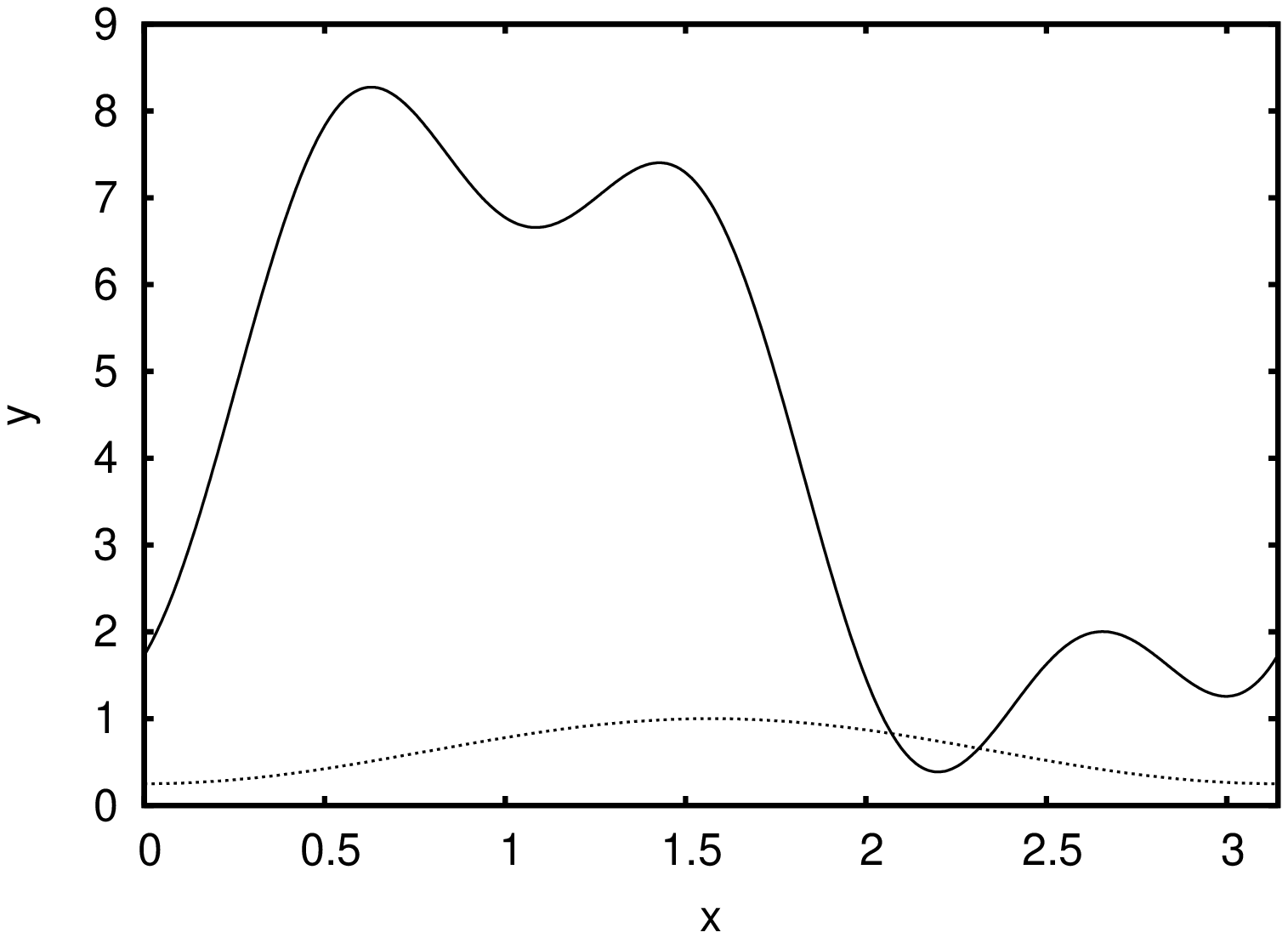}
\caption{The current matrix element $|\langle E|[{\bf H},x]|-E\rangle|^2$
as a function of the angle $\varphi$ between the wave vector ${\bf k}$ and the 
direction of the current operator.
The left panel is the matrix element for $E=2$ without gap ($m=0$) and the right panel
is the matrix element with gap ($\Delta=2m=2$). Full (dashed) curves are for BLG (MLG).
}
\label{current}
\end{center}
\end{figure}

{\it current matrix elements:}
The commutator in the current operator is for Dirac fermions the Pauli matrix
$\sigma_\mu$: $[{\bf H}_M,r_\mu]=i\sigma_\mu$ (MLG) and for BLG
in Fourier representation
\[
[{\bf H}_B,x]=  i\frac{\partial {\bf H}_B}{\partial k_x}
=2i(k_x\sigma_1+k_y\sigma_2)
\ .
\]
Then for the current matrix element for MLG we obtain
\beq
|\langle E|\sigma_1|-E\rangle|^2 =\frac{m^2+k_y^2}{E^2}
\label{current2m}
\eeq
which after angular integration yields
\beq
\int_0^{2\pi}|\langle E|\sigma_1|-E\rangle|^2 d\varphi 
=\pi\frac{2m^2+k^2}{E^2}
=\pi\frac{m^2+E^2}{E^2}
\ .
\label{memlg}
\eeq
For BLG (with $k^2=\sqrt{\omega^2/4-m^2}$, $k_x=k\cos\varphi$ and $k_y=k\sin\varphi$) we have
\beq
|\langle E|k_x\sigma_1+k_y\sigma_2|-E\rangle|^2
=k^2\Big[\frac{m^2}{E^2}+4\frac{k^4}{E^2}
(\cos^2\varphi-\sin^2\varphi)\cos^2\varphi\sin^2\varphi
+\frac{k^4}{E^2}\sin^2\varphi
+\frac{2m}{E}\cos\varphi\sin\varphi
\Big]
\label{current2b}
\eeq
%\[
%\frac{|\langle E|k_x\sigma_1+k_y\sigma_2|-E\rangle|^2}
%{\langle E|E\rangle \langle -E|-E\rangle}
%=\sqrt{E^2-m^2}\Big[\frac{m^2}{E^2}+4\frac{(E^2-m^2)}{E^2}
%(\cos^2\varphi-\sin^2\varphi)\cos^2\varphi\sin^2\varphi
%\]
%\beq
%+\frac{(E^2-m^2)}{E^2}\sin^2\varphi
%+\frac{2m}{E}\cos\varphi\sin\varphi
%\Big]
%\label{current2b}
%\eeq
and after the angular integration
\beq
\int_0^{2\pi}|\langle E|k_x\sigma_1+k_y\sigma_2|-E\rangle|^2 d\varphi 
=\pi k^2\frac{m^2+E^2}{E^2}
\ .
\label{meblg}
\eeq
This is valid only for $\omega^2\ge 4m^2=\Delta^2$.
As an example, these current matrix elements are plotted for $E=2$ in Fig. \ref{current} with and
without gap.

%\section{conductivity}

{\it conductivity:}
Now we insert the results of Eq. (\ref{memlg}) into the Kubo formula
Eq. (\ref{kubo0}) and obtain for MLG
\beq
\sigma_{MLG}'=\frac{\pi e^2}{8h}\left[1+\frac{\Delta^2}{\omega^2}\right]\Theta(\omega^2-\Delta^2)
[f_\beta(E_F+\omega/2)-f_\beta(E_F-\omega/2)]
\ .
\label{ocond2}
\eeq
Inserting  Eq. (\ref{meblg}) into the Kubo formula gives exactly twice the
conductivity of MLG: $\sigma_{BLG}'=2\sigma_{MLG}'$.
%\[
%\sigma_{xx}'=\frac{\pi e^2}{4h}\left[1+4\frac{m^2}{\omega^2}\right]\Theta(\omega^2-4m^2)
%[f_\beta(E_F+\omega/2)-f_\beta(E_F-\omega/2)]
%\ .
%\]
Thus the conductivities of MLG and BLG agree up to a factor 2. 
The additive correction due to the gap parameter $\Delta^2$ decays like $\omega^{-2}$.
As an example, the behavior of the conductivity versus $\omega/k_BT$ is plotted in Fig. \ref{conductivity}.

\section{Discussion}

%{\it discussion:}
The conductivity in Eq. (\ref{ocond2}) obeys a scaling relation of the conductivity:
\[
\sigma_{xx}'(\beta,\Delta, E_F,\omega)=\sigma_{xx}'(\beta \Delta,\beta E_F,\beta\omega)
=\sigma_{xx}'(\Delta/\omega,\beta E_F,\beta\omega)
\ ,
\]
i.e. the conductivity depends only on three independent parameters. A similar scaling
relation exists also for the conductivity in the presence of a scattering rate 
\cite{ziegler07}. This implies that a reduction of temperature is equivalent to a
simultaneous increase of frequency, gap and Fermi energy.

\begin{figure}[t]
\psfrag{x}{$\omega/k_B T$}
\psfrag{y}{$\sigma_{xx}'\ \ [e^2/h]$}
\begin{center}
%\psfrag{H2}{${\cal H}_2$}
%\psfrag{H4}{${\cal H}_4$}
%\includegraphics[width=7cm,height=7cm]{ocond2.eps}
\includegraphics[width=7cm,height=7cm]{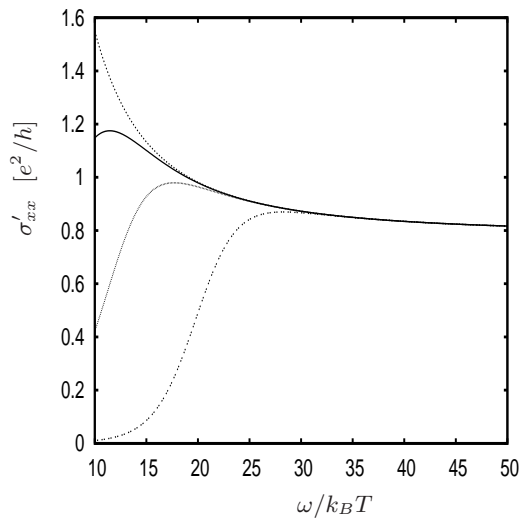}
\caption{Optical conductivity of BLG as a function of $\beta \omega$ for $\beta m=5$
and $\beta E_F=0,4,6,10$ with decreasing conductivity. These curves have to be multiplied
by 1/2 to get the corresponding values for MLG.
}
\label{conductivity}
\end{center}
\end{figure}

The optical conductivity vanishes for photon energies less than the gap.
It jumps to some finite value when the photon energy exceeds the gap energy,
where the size of the jump depends on the Fermi energy (cf. Fig. \ref{conductivity}).
The maximal jump appears at $E_F=0$. This allows us to measure the gap by
measuring the jump of the optical conductivity. Such a measurement can be
performed as an optical transmission experiment \cite{kuzmenko08,nair08} because
the optical conductivity is related to the transmittance $T$ through the relation 
\[
T=\frac{1}{1+2\pi\sigma_{xx}'/c}
\ .
\]  
The parameter $c$ is the speed of light. Then graphene is completely transparent for
photonic energies up to the gap energy. This could be used to filter light
with frequencies higher than the one given by the gap energy.  

Our model assumption of taking into account only the low-energy bands of
BLG restricts the photon energies $\hbar \omega$ to less than $0.8$eV,
which is the gap between the low-energy and the high-energy bands in BLG
\cite{abergel07,nicol08}. This restriction also avoids stronger
deviations from the Dirac theory of MLG and van Hove singularities. 
Taking into account high-energy
bands does not change this picture qualitatively, since it would
lead to additional jumps of the optical conductivities as soon
as the photon energies exceed the gap energies. A van Hove singularity
would appear as an additional peak.
The effect of the gap is a global enhancement by $m^2$, where
for BLG the situation is more complex than for MLG.

In conclusion, focusing on the low-energy dispersion of gapped monolayer
and bilayer graphene, we have found simple expressions for the
optical conductivities. They agree for all parameters up to a factor 2.
This is remarkable because the current-matrix elements of both graphene
systems are rather different.

\end{document}